# Multi-Energy Radiography Against Terrorism. Theory and Experiments

Sergei V. Naydenov, Vladimir D. Ryzhikov, Craig F. Smith, Dennis Wood

*Abstract*--**Multi-energy radiography is a new direction in non-destructive testing. Its specific feature is separate detection of penetrating radiation in several energy channels. Multi-energy radiography allows quantitative determination of the atomic composition of objects. This is its principal advantage over conventional radiography. In particular, dual-energy radiography allows determination of the effective atomic number of a material with an accuracy of up to 80-90%. Development of three-energy radiography and radiography of higher multiplicity makes it possible to further improve the reconstruction of an object's chemical composition. This presents the possibility, for example, of detection of explosives and other illegal objects in luggage with a reliability approaching 100%. These developments can find application not only in anti-terrorist activities, but also in areas such as industrial testing and nuclear medicine.**

## I. INTRODUCTION

Among the recent technological novelties of the 21st century, one should note the revolutionary replacement of conventional film radiography by digital technologies. In digital radiography, the test object is inspected by X-ray or gamma-radiation, and the object image is reconstructed in digital form in near real time, using linear and/or planar solid-state scintillation detectors. These technological changes have dramatically influenced the methods used for recording, storing and processing useful information. An important case in point is the new multi-energy approach presented in this paper. The two-energy version was first used in medical studies, see [1]-[3], etc. An object (i.e., a material item or a patient) is irradiated, and the radiation that passes through the object is recorded separately at several characteristic discrete energies and/or in several separated energy ranges.

Two-energy (dual-energy) radiography is the simplest variant of multi-energy radiography. However, its practical realization present a rather complex technical problem. In this paper, we consider the general principles of multi-energy radiography, as well as its application to anti-terrorist activities. The multi-energy approach allows one to improve the inspection efficiency by an order of magnitude in contrast to single energy methods. This improvement is related to the ability to reconstruct the atomic structure of materials. In conventional radiography methods, it is only the spatial characteristics of the inspected objects that can be reconstructed.

## II. PHYSICAL PRINCIPLES

Let us first consider the unique feature of two-energy radiography. Detectors in such systems record the transmitted (attenuated) ionizing radiation not in just one, but in two separate ranges of the energy spectrum. Fig. 1 presents the general scheme of multi-energy radiography in comparison with the conventional, single energy approach. Each channel of detection corresponds to one of the characteristic energies of radiation. Methods of energy separation depend upon the detector design and the choice of radiation sources. In a classical design with X-ray tubes, metal filters are used [4]. In new developments, the role of a filter may be played by a low-energy detector array (e.g., based on zinc selenide).

The better the separation of the energy ranges is, the higher is the inspection efficiency and the quality of the object reconstruction. An ideal choice would be the use of monochromatic X-ray sources. However, industrial development of such sources (e.g., X-ray lasers) is only at its preliminary stages. Practically-available X-ray tubes yield continuous radiation spectra. With such a distributed spectral source, energy separation is achieved by selecting the most efficient scintillators for detection of radiation in different spectral ranges, as well as by design optimization of the detectors and their arrays.

From multi-energy imaging data, the physical parameters of the object and the technical parameters under control are reconstructed with the help of special algorithms. To

Manuscript received November 10, 2004. The work was supported in part by the U.S. Civilian Research and Development Foundation for the Independent States of the Former Soviet Union under CRDF Project No. UE2-2484-KH-02.

Sergei V. Naydenov and Vladimir D. Ryzhikov are with the Institute of Single Crystals of the National Academy of Science (NAS) of Ukraine, Kharkov, 61001 Ukraine (telephone: +380-57-330-8316, e-mail: naydenov@isc.kharkov.ua, ryzhikov@isc.kharkov.com ).

Craig F. Smith is with Lawrence Livermore National Laboratory, Livermore, CA 94550, USA (telephone: 1-925-423-1772, e-mail: smith94@llnl.gov ).

Dennis Wood is with Rhyolite Technology Group, Inc., 1191 St. Charles Court, Los Altos, CA 94022, USA (telephone: 1-650-248-1950, e-mail: dennis@denniswood.com ).

reconstruct the material structure, we have developed a new physical model and appropriate algorithms for solution of the corresponding inverse problems. A detailed theoretical analysis was described in our previous papers [5]-[6].

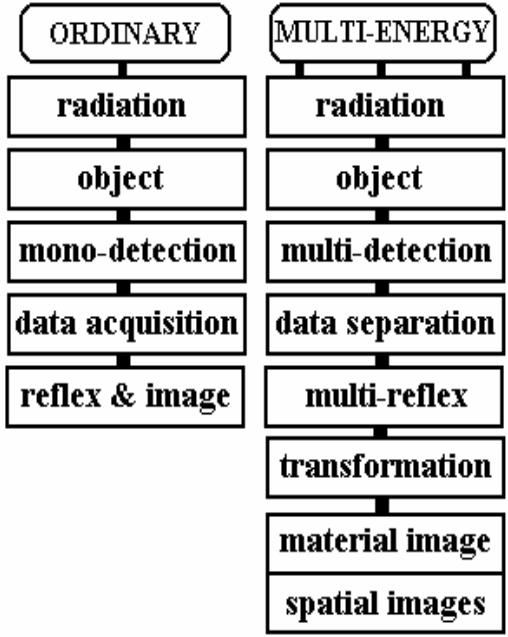

Fig. 1(a). Comparative scheme of ordinary and multi-energy radiography. The studied object in both cases is inspected by penetrating radiation (X-ray or gamma quanta). However, in multi-energy radiography, the radiation that passes through the object is separately recorded not in just one, but in several output channels. A set of digitized radiographic reflexes (a multi-component vector image) allows, after transformation according to the chosen radiography algorithm, reconstruction of information on the physical structure of the material, with separation of spatial images of specific parts of the object having different characteristics of absorption.

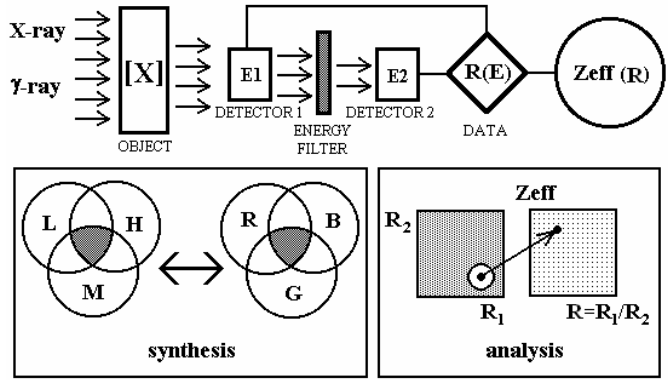

Fig. 1(b). Reconstruction of $Z_{eff}$ in dual-energy radiography. The conventional scheme consists of obtaining a best fit from the combination of basic atomic mass characteristics: L for "light", M for "middle" and H for "heavy" atomic mass. "Black-and-white" synthesis corresponds to the results obtainable from two-energy radiography; the "three-color" scheme (R – red, G – green, and B – blue) corresponds to 3–radiography, etc. Our method (analysis) provides the basis for the unambiguous reconstruction of $Z_{eff}$.

Our approach is marked by a peculiar feature. For the most accurate reconstruction of material structure, we require not only good energy separation (as in other multi-energy methods); we also require that the energy mixing spans multiple absorption channels representing the different fundamental processes of radiation scattering: i.e., the photo and Compton effects (at 80-200 keV), or pair production and Compton effects (at 2-8 MeV). The competition between the channels does not degrade inspection efficiency, but, on the contrary, enhances it. Thus, the absorption channels should be mixed, while the recording channels should be separated. This requires a subtle matching of the detection system and the characteristic radiation energies. Our theoretical and experimental studies in this direction indicate a possibility of determining the atomic composition of inspected materials with an accuracy of up to 80-90%.

### III. EFFECTIVE ATOMIC NUMBER

Among the most important material structure parameters, one should specially note the effective atomic number, $Z_{eff}$. In fact, it correlates well with the chemical composition of a material. Higher values ($Z_{eff} \geq 20$) correspond to inorganic compounds and metals, and lower values ($Z_{eff} \leq 10$) correspond to organic substances. For many applications, e.g., for radioisotope monitoring of the content of nuclear fuel elements or radioactive wastes, this parameter has a decisive significance. Analysis of $Z_{eff}$ is also necessary for geological studies of ore and mineral composition, search for new sources of fossil fuels, structure monitoring of composite materials, separate diagnostics of soft and bone tissues in medical radiology, determination of calcium content in tests for osteoporosis, etc.

Normally, in radiography the material is irradiated in the ranges corresponding to the photo and Compton effects (from 20-50 keV to 200-400 keV). The effective atomic number of a compound of known chemical formula is determined from the relationship

$$Z_{eff} = \left[ \sum_{k=1}^{N} a_k A_k Z_k^4 \Big/ \sum_{k=1}^{N} a_k A_k Z_k \right]^{1/3}, \quad (1)$$

where $A_k$ and $Z_k$ are the atomic mass and atomic number of simple elements, respectively; $N$ is the total number of simple elements; and the values $a_k$ represent the relative atomic (molar) concentrations, i.e., the number of atoms of each kind in one molecule.

### IV. RADIOGRAPHIC LAW

We have proposed a new method for determination of $Z_{eff}$. In traditional methods, the effective atomic number is determined as a superposition of several reference values (see Fig 1(b)). The weights in this superposition determine the relative content of light, heavy and/or medium "basic" materials. Therefore, the weights depend upon the choice of the basis. Inappropriate choice of the basis leads to errors. For example, a thin sample of a material with a large atomic number can be perceived as a thick sample of a material with a lower atomic number. With increased degree of energy multiplicity, the errors that arise using the basic materials method decrease. But even in these cases, errors can reach tens of percents. Thus, a distinction can be discerned between materials such as iron and wood, or iron and light alloys. But it is not possible to distinguish gold from lead, iron from nickel, calcium minerals from sand, explosives from organics, etc.

Our theoretical results have allowed us to establish a universal radiographic law to improve the ability to identify material composition. Here, the concept of relative reflex plays the principal role. For objects of arbitrary type and composition, there is a universal relationship between this reflex and the effective atomic number. In luggage inspection, one should work at the energies where the photo and Compton effects are predominant. For inspection of ship containers, high energies provided by linear accelerators can be required, at which the pair production and Compton effects are predominant.

In theory, two-energy radiographic reflex is defined as the ratio of single-energy reflexes obtained separately from each of the assembly detectors:

$$R = R_1/R_2 \; ; \tag{2}$$

$$R_i = R(E_i) = \ln\left[V_{\text{back}}(E_i)/V(E_i)\right] \; . \tag{3}$$

Here, $V_{\text{back}}$ is the detector signal in the background mode, when the inspected object is absent; $V$ is the signal recorded by the detector under irradiation of an object. Depending upon the chosen detection mode, the digital signal $V$ is proportional to the counting rate of the pulse analyzer at the output of the spectrometric circuit (counting or spectrometric detection mode) or to the measured current at the output of the electronic amplifier circuit (current mode). The output signal can be normalized and then digitalized by any convenient method. The monitoring results do not depend upon normalization and the choice of measurement units, as it follows from (3). However, for assemblies (matrices, arrays) of detectors, the normalization of signals should be harmonized, i.e., they should be adjusted so that output signals are the same for each separate element of the system in both measurements – in the background mode and in the presence of an object.

The essential role of the relative reflex $R$ is based on the fact that it does not depend upon density $\rho$ or thickness $d$ of the inspected material, but only upon the effective atomic number. Within the framework of conventional models of exponential absorption of ionizing radiation, it can be shown that:

$$R = \mu_m(E_1)/\mu_m(E_2) = f(Z_{\text{eff}}) \; , \tag{4}$$

where $\mu_m = \mu_m(E) = \mu(E)/\rho$ is the mass coefficient of linear attenuation for radiation with energy $E$ in the inspected material. Hence, it follows that $R \neq f(\rho, d)$. It is important that there exists an unambiguous relationship between the reflex $R$ and the effective atomic number $Z_{\text{eff}}$. This relationship is expressed by the following theoretical law:

$$Z_{\text{eff}} = \left[(c_1 R + c_2)/(c_3 R + c_4)\right]^{1/(p-1)} \; , \tag{5}$$

where $c_1, c_2, c_3, c_4$ and $p$ are fixed calibration parameters.

Such a relationship follows from the solution of the corresponding inverse problem of reconstruction of the material parameters from multi-energy data resulting from X-ray inspection, accounting for at least two different absorption mechanisms with different dependence upon the effective atomic number. In other words, the characteristic law (5) is necessarily valid, e.g., at medium energies from 50-100 keV to 200-300 keV, where there are contributions from both the photo effect (with mass attenuation coefficient $\mu_{m,\text{photo}} \propto Z_{\text{eff}}^4$) and the Compton effect; or in the energy range of linear accelerators from 1-2 MeV to 5-7 MeV, where the pair production effect ($\mu_{m,\text{pairs}} \propto Z_{\text{eff}}^2$) and Compton scattering are predominant. At very low energies, when there is no pair production and the Compton contribution is weak, or at very high energies, when photo and Compton effects are strongly suppressed, deviations from the predicted law can be observed. Within the considered inverse problem, $Z_{\text{eff}}$ is determined by only one measured parameter, namely, by the said relative radiographic reflex $R$. The density of the inspected unknown material is a function of not one, but of two variables $\rho = \rho(R_1, R_2)$, demonstrating a more complex dependence upon single-channel reflexes $R_1$ and $R_2$.

The calibration constants $c_1, c_2, c_3, c_4$ in (5) depend upon the chosen monitoring scheme, energy values used in the radiation source, characteristics of the receiving-detecting system, etc., but not upon properties of the inspected material. These constants are determined from a series of test measurements using objects of known composition and geometry. The power index $p$ depends upon which absorption mechanism is predominant in the inspected material. For the medium energy range, where the photo effect is predominant in combination with the Compton effect, one can set $p = 4$. At high energies, when the pair production effect predominates, one should set $p = 2$.

The proposed law is applicable for all energy values (low, medium, high). This is confirmed by the available experimental data. In Fig. 2, for verification of the proposed law, data are presented on radiation absorption following two-energy irradiation of simple substances. Known experimental data taken from different sources were used [7]-[8] describing absorption of X-ray radiation at different energies. In Fig. 2(a),

X-ray inspection data are presented for test samples at 127 keV and 62 keV, and in Fig. 2(b) – at 127 keV and 95 keV, where the recorded energies differ strongly (by about two times) or slightly (by about 30%). Experimental data on absorption in different materials at two different energies were used for calculation of relative radiographic reflexes. A theoretical plot was constructed from (5) basing on calibration data for only three chosen materials (this is sufficient, since one of the four constants $c_1, c_2, c_3, c_4$ can be expressed in terms of the other three). These substances can be chosen arbitrarily, but should have $Z_{eff}$ values that are close to each other. For such calibration, we have chosen two elements (carbon (C) and lead (Pb)) at the edges of the radiographic "spectrum" shown in Fig. 2, and one element (iron (Fe)) in the middle of it. With such choices, the experimental data for other elements fit the predicted theoretical curve with high accuracy. It should be noted that, for different radiographic configurations accounting for the specified energy choice, as well as for different irradiation and detection conditions, this curve will vary, since it depends on the calibration parameters of the monitoring. But its general character and its linear fractional dependence (5) will remain the same.

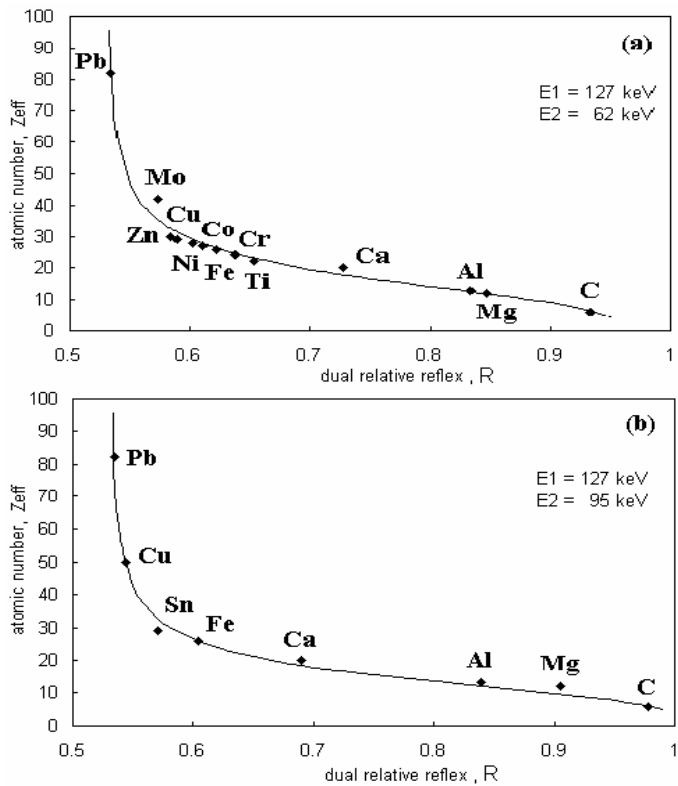

Fig. 2. Dependence of the effective atomic number upon radiography reflexes in the regions of the photo effect and Compton scattering. The theoretical dependence (solid line) and the experimental points for materials of known composition are indicated. Geometry (dimensions) of samples is arbitrary.

As shown in Fig. 2, the theory appears to be in good agreement with experiments. In addition to the above-presented results, we have also found good correspondence between theory and experiment (errors not exceeding 10%) in other ranges of the energy spectrum – in the regions where Compton scattering or pair production are predominant, as well as at different ratios of the two chosen radiation energies (detection energies) $E_1$ and $E_2$ (from 2:1 to 4:3). Our analysis confirms the universal character of the dependence shown in (5). In our further studies, we plan to carry out similar comparative assessments of complex substances and composite materials (mixtures, solutions, superimposed objects, etc.).

V. SEPARATION OF NON-ORGANICS FROM ORGANICS

In standard dual-energy custom inspection systems such as those produced with the participation of the Institute of Single Crystals (refer to Fig. 3), we have achieved good distinction between substances with similar density but significantly different atomic number. Fig. 4 demonstrates this achievement showing a picture of a set of inspected objects together with the images obtained in the X-ray inspection.

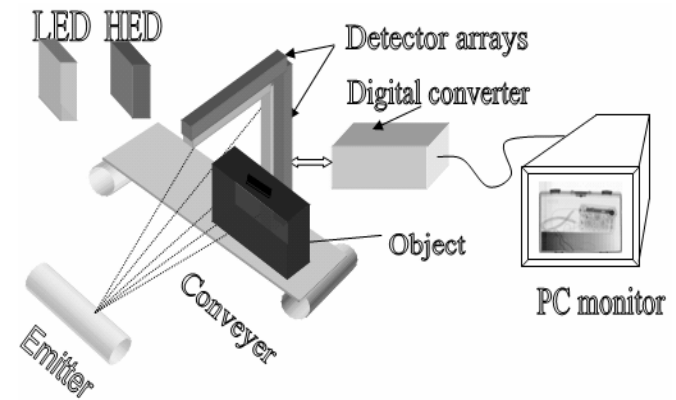

Fig. 3. With the scintillator-photodiode receiving circuit, we used a 16-channel photodiode (FD-321 produced by CCB "Ritm" and the Institute of Microdevices (Ukraine)) using a step size of 1.65mm, and a total of 128 channels. We also used a 32-channel photodiode (PD) of the same producer, with a step size of 0.8mm, and a total of 256 channels. The scintillators used were CsI(Tl), CdWO4 and ZnSe(Te) crystals produced by the Institute of Single Crystals (Ukraine).

Accounting for the trend in modern digital radiography toward use of two-energy detector arrays, the combination of detector crystals based on ZnSe(Te) and CsI(Tl)/CWO represents a significant advance. The effective atomic number of ZnSe is the same as that of copper, which is commonly used as a filter in high-energy arrays. Therefore, if a detector with ZnSe(Te) is used, it also can serve as a filter when placed before the high-energy array. In so doing, one can simplify the design and improve the technical characteristics of the detecting circuit as a whole. A unique combination of properties that characterize the ZnSe-based scintillator (high light output, fast response, radiation stability, rather low effective atomic number together with sufficiently high density) makes this material the best among known scintillators for the low-energy detector. The combination of ZnSe(Te)/CsI(Tl) crystals in the two-energy detector array has substantially improved the sensitivity of the equipment designed for detection of organic inclusions.

*A. MER with scintillator-photodiode detectors on base ZnSe*

Let us briefly consider the reasons why the use of ZnSe scintillators in low-energy detectors for two-energy radiography leads to substantial improvement in quality and accuracy of monitoring. In Table 1, comparative characteristics are presented for three promising scintillators used in multi-energy radiography. Analysis of these data shows that ZnSe(Te) scintillators should be preferable for detection in the low-energy spectral region; CsI(Tl) – in the region of medium energies; and scintillators such as CWO (Cadmium Tungstate, $CdWO_4$) – for high energies. Heavy oxide scintillators (CWO or BGO (Bismuth Germanate, $Bi_4Ge_3O_{12}$)), with their high atomic numbers, will be used in future for three-energy radiography.

To increase the contrast sensitivity of dual-energy radiography and ensure the clear discrimination of images related to organic and inorganic components of the object, the detection energies (channels) should be separated as clearly as possible. This means that in the two-energy assembly, the low-energy detector (LED) should attenuate (while detecting) the low-energy part of the X-ray radiation spectrum used for inspecting the object. Similar properties are expected from the high-energy detector (HED) for the remaining high-energy part of the spectrum. To ensure this, high-energy radiation should freely pass through the LED without substantial attenuation. In a good radiographic installation, two apparently contradictory conditions should be met: 1) there should be maximum attenuation in the LED of the low-energy radiation passing through the object – it should not reach the HED; 2) there should be maximum transmission of the high-energy radiation through the LED and other parts of the receiving-detecting circuit – it should be attenuated only in the HED. The use of metal filters allows one to solve only the first of these problems. For the second problem, filters are of no use and are even disadvantageous, because some of high-energy gamma-quanta are absorbed by the filter. There exists another efficient way – to use the low-energy scintillator itself as a filter for energy separation. Such a scintillator should possess the following unique properties – high light output and small radiation length with respect to low-energy photons, as well as good transparency for high-energy photons, i.e., low effective atomic number. Our theoretical and practical studies have shown that the most suitable material for those purposes is zinc selenide and semiconductor compounds based on it.

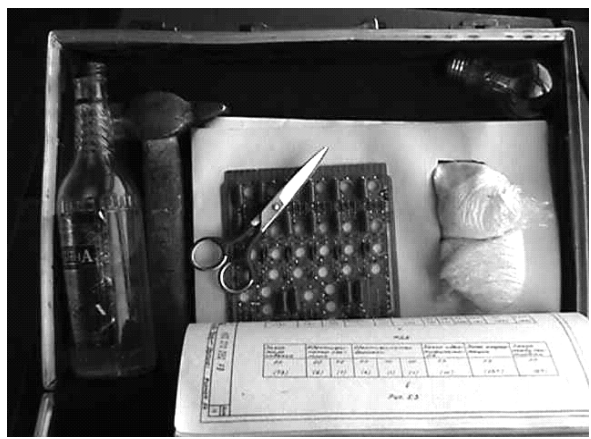

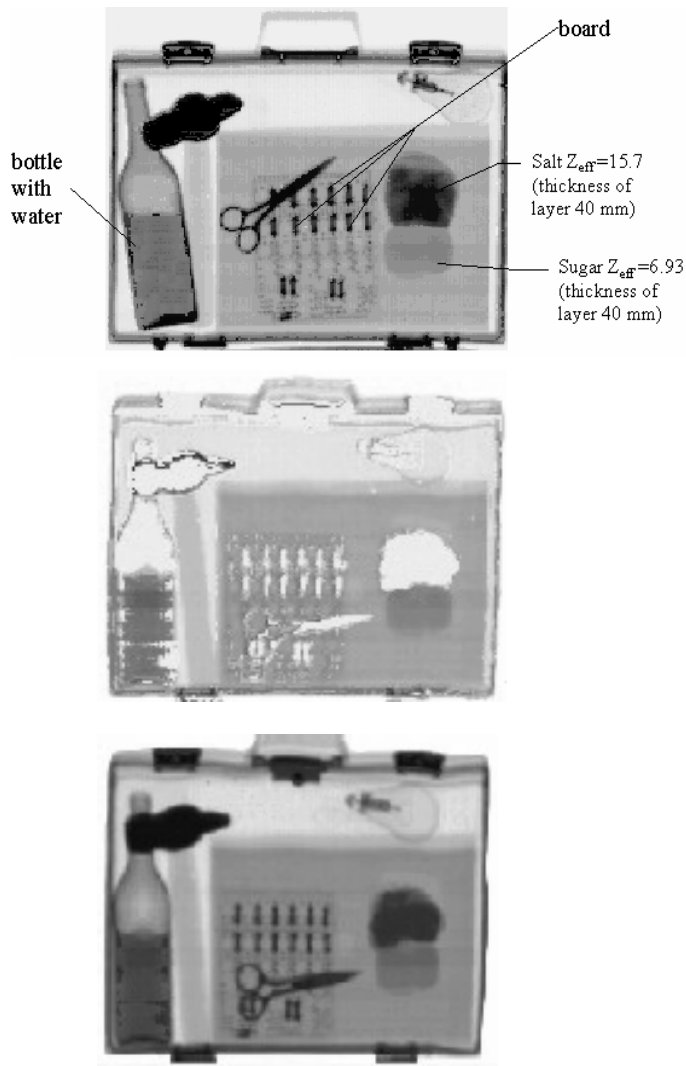

Fig. 4. Snapshot of inspected object and its X-ray shadow image views. Included are the real image and separate images showing distinction of organic and non-organic components (see from top to down). On the second from the bottom picture, the hammer head (iron) is excluded from the image, its contour is an error of the image reconstruction program due to parasitic scattering of X-ray radiation at the edges of solid objects made of materials with high $Z_{eff}$. Wood (the handle of the hammer) is distinguished on the original computer color images of the object. After transformation of images to black and white, this distinction became barely noticeable.

TABLE I
COMPARATIVE CHARACTERISTICS OF SCINTILLATORS

| Parameter | Scintillator | | |
|---|---|---|---|
| | ZnSe(Te) | CsI(Tl) | CWO |
| Density, g/cm$^3$ | 5.42 | 4.51 | 7.99 |
| Effective atomic number | 33 | 54 | 65 |
| Hygroscopic | no | no | no |
| Luminescence maximum wavelength, nm | 630-640 | 550 | 490 |
| Afterglow:  in 5 ms, %  in 20 ms, % | < 0.2  < 0.05 | 0.3-2.0 | < 0.05  ≤ 0.02 |
| Intrinsic radiation absorption coefficient, cm$^{-1}$ | 0.05-0.2 | < 0.05 | < 0.05 |
| Light output with respect to CsI(Tl) at low energies, % | 100-140 | 100 | 30 |
| Decay time, μs | 50-150 | 1 | 10-15 |
| Coefficient of spectral matching with PD, % | 0.92 | 0.75 | N/A |
| Light yield, up to photon/MeV | 7.5·10$^4$ | 5.5·10$^4$ | 1.3·10$^4$ |
| Radiation stability, Rad | 5·10$^8$ | 5·10$^4$ | 1·10$^8$ |

Fig. 5 shows measurement data indicating the sensitivity (response) of different detection systems to X-ray radiation of different energies. In the low-energy range, ZnSe has an even higher light output than the well-known crystal CsI(Tl), being slightly inferior in this respect at higher energies.

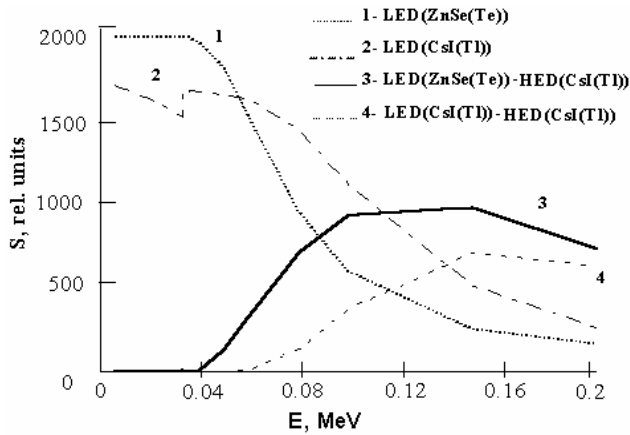

Fig. 5. Sensitivity (in relative units) of individual detectors and their assemblies versus energy of applied X-ray radiation. Plots 1 and 2 were recorded for low-energy detectors (LED) on the basis of scintillators ZnSe(Te) and CsI(Tl). Plot 3 shows the level of the recorded signal from radiation passed through a combined assembly LED[ZnSe(Te)]-HED[CsI(Tl)] comprising different scintillators. Plot 4 shows the sensitivity of an assembly made on the basis of two CsI(Tl) scintillators having different thicknesses (the LED crystal is about four times thinner than the HED one). Scintillator thickness was optimized to ensure full absorption of a selected part of the energy spectrum, i.e., the low- or high-energy component.

The sensitivity of the LED[ZnSe(Te)]-HED[CsI(Tl)] assembly, where zinc selenide plays two roles, detection element and energy filter, is nearly twice as high as that of the LED[CsI(Tl)]-HED[CsI(Tl)] assembly made of similar scintillators of different thickness. The variant LED[ZnSe(Te)]-HED[ZnSe(Te)] (not shown in Fig. 5) will be even worse because ZnSe has lower light output as a HED. The worst of all would obviously be the LED[CsI(Tl)]-HED[ZnSe(Te)] assembly. As it can be seen from Fig. 5, in the working range (about 100-120 keV), the sensitivity of the LED[ZnSe(Te)]-HED[CsI(Tl)] assembly is only weakly dependent upon energy. This is also important, because it provides efficient avoidance of fluctuations related to the non-monochromatic character of the source radiation. The energy separation becomes optimal in such a system. Therefore, one should expect high accuracy in determination of the effective atomic number of the inspected material. In our experiments, we used a LED[ZnSe(Te)]-HED[CsI(Tl)] assembly with thickness $d_1 = 0.6$ mm for ZnSe(Te) and $d_2 = 4$ mm for CsI(Tl), which appeared to be optimal upon irradiation of the objects by X-ray source with voltage $U_a = 140$ kV on the tungsten anode.

*B. Medical applications of MER*

In this paper, we concentrated on the development of multi-energy radiography for anti-terrorist activities. However, possible MER applications in medicine should also be noted. Advanced medical methods for the distinction between osseous and soft tissues are based on distinction between their effective atomic numbers. The two-energy approach is known to provide, in comparison with conventional radiography, substantially higher contrast sensitivity in separation of radiographic images of osseous and soft tissues, see [1]-[3] etc. Two-energy radiography obtains distinction of the object images by their effective atomic number and not by their radiographic density. This ensures accuracy that is an order of magnitude higher compared with ordinary radiography, the contrast sensitivity of which is often not sufficient for distinction between superimposed hard and soft tissues. Further, multi-energy methods may be capable of discriminating between different types of non-osseous tissue.

This quality improvement of medical diagnostics is directly related to the accuracy of radiographic determination of the effective atomic number and density of different tissues and organs. Our developments are aimed at distinction between materials with effective atomic numbers differing by unities and tenths of unity, such objects being of relevance for modern radiology. Examples are the quantitative determination of calcium depletion in bones due to osteoporosis or the diagnostic analysis of plaques. Hard plaques, related to calcium sedimentation on blood vessel walls, can be detected by modern methods, including computer tomography. Identification of soft plaques, with their chemical composition being close to that of blood, has been practically impossible. Development of our new methods for direct quantitative reconstruction of the variable profile of $Z_{eff}$ and density $\rho$ in such biological objects (e.g., in blood clots) can be of substantial importance.

## VI. DETECTION OF EXPLOSIVES AND ILLEGAL OBJECTS

With our standard 2-energy customs inspection instruments, "Polyscan" [9] (series Polyscan-4, 5, and 6), we have no possibility to distinguish between substances with close atomic numbers. However, with our experimental instruments, which use a new – unique and exclusive – scintillator based on ZnSe(Te) having very high sensitivity for low energy and simultaneously sufficient transparency for high energy X-rays, we can distinguish between substances with close atomic numbers (see Figs. 6-7): soap ($Z_{eff}$ =5.9-6.5) or wood ($Z_{eff}$ =6.5-7.3) and an explosives ($Z_{eff}$ =7.0-7.7). As it is also shown in Fig. 7 we can see also sufficient spatial resolution for details of illegal objects. This result will be, in our opinion, of a great importance for future application in security systems.

### A. Physical Accuracy

The proposed method is free from the drawbacks related to the choice of basic materials. The theory gives a proportional relationship between the sensitivity of our method and the thickness sensitivity:

$$\Delta Z/Z \propto 2\,(\Delta d/d) \ . \qquad (6)$$

The term ($\Delta d/d$) is the sensitivity of an ordinary radiography; normally it is about several percent. The factor of 2 in (6) is related to the presence of two detectors for registration of two energies. The theoretical limit of accuracy would be about 80-90% if the spatial resolution of the detectors could be made better than 4 line pairs/mm.

The universal character of the radiographic law was established and verified for ideal conditions, i.e., for the case of full energy separation and detection of two radiation channels (a source with two well-separated monochromatic lines in the spectrum), for uniform objects of simple substances, in the absence of noise, etc. In principle, the determination error for the effective atomic number can be reduced to several percent. This means that it is possible to distinguish between materials with effective atomic numbers that differ by as little as 1-2 units. This ability is demonstrated in Fig. 2, where materials of close effective atomic numbers are distinguished (Ti, Cr, Fe, Co, Ni, Cu, Zn). However, in real conditions, most error sources cannot be neglected, and the monitoring accuracy obtained can be lower by several times. Major factors affecting the monitoring accuracy are: 1) the non-monochromatic character, non-uniformity and instability of the X-ray emitter spectrum; 2) incomplete separation in detection of low- and high-energy components due to non-optimized detector assembly configuration; 3) statistical and other fluctuations, including poor energy resolution of detectors, circuit noise and dynamic noise due to slow system response on the real time scale (especially for moving objects); 4) scattered radiation inside a non-uniform object and deterioration of spatial resolution; 5) coincidence or overlapping of constituent parts of a complex object (e.g., when inorganics shield organics, or different organic materials are mixed; and 6) hardware or software drawbacks in digitalization, 3D-imaging and separation of images of different components of the object, including organic and inorganic. By removing or minimizing the effects of these factors, we expect 90% accuracy in discrimination of materials by their effective atomic number.

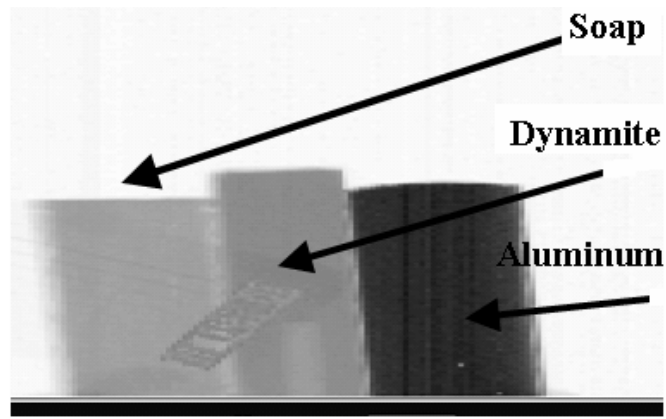

Fig. 6. Identification of different substances using the dual-energy method. The organic materials shown (soap and explosive) differ in their effective atomic number by not more than 10-15%. To compare the MER contrast sensitivity for different materials, in the right-hand side of the image is presented a piece of inorganic aluminum, the effective atomic number of which is nearly two times higher than that of the organic objects.

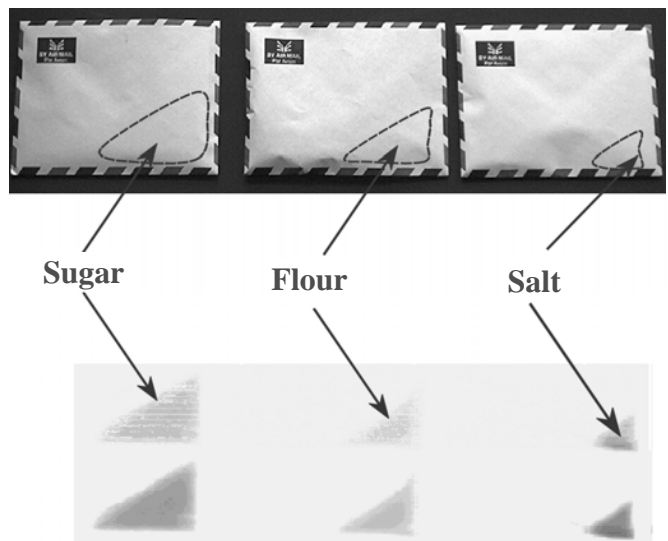

Fig. 7. Examination of powder materials in postal parcels. The images below the envelopes are the low-energy and high-energy images received by LED (upper) and HED (lower) detectors of the inspection system.

*B. Actual Problems*

The accuracy of determination of the atomic number described in our work is obtained in the approximation of monochromatic radiation. In real devices, X-ray tubes with broad continuous spectra are used. Thus, detectors are required that are selective for energies in specified energy spectrum ranges, which increases the cost of the equipment. An optimum balance between price and quality is yet to be found.

For detection of very small quantities of dangerous inclusions, high detector sensitivity is required, which requires a larger detector volume. At the same time, the requirement of high spatial resolution requires small output apertures. This is needed to exclude the effects of scattered radiation (the contribution of scattering is especially large for inclusions with high $Z_{eff}$). Again – a reasonable compromise is needed.

Our analysis has shown that multi-energy radiography can be used for determination of concentrations of simple elements in complex substances. It is essential that the number of components correspond to the degree of energy multiplicity. For monitoring of two-component systems, dual radiography (two-energy radiography) is sufficient. Explosives are generally characterized by elevated content of nitrogen and oxygen. For a more precise identification of an organic compound, three elements should be determined (carbon, nitrogen, oxygen) or all four (hydrogen, carbon, nitrogen, oxygen). Such problems of quantitative diagnostics should be dealt with through technical development of, respectively, 3-energy and 4-energy radiography.

## VII. SUMMARY

In this work, we have shown theoretically and confirmed experimentally that it is possible, within the framework of two-energy radiography, to separate substances with a rather small (10-20%) difference in effective atomic numbers. This capability is immediately applicable to security inspection equipment (i.e., for airports, customs services, etc.), as well as to medical diagnostics and non-destructive testing.

The practical results described in this work have been achieved largely by using an original scintillator based on ZnSe(Te) in the low-energy detection circuit. This scintillator is highly efficient in the low-energy range (30-60 keV) and has low sensitivity in the high-energy range (above 100 keV).

In general, these results indicate a promising new direction in research and application of radiographic spectroscopy of materials. In multi-energy radiography, it becomes possible to distinguish separate components of otherwise physically non-separable composite and multilayer objects. It can be used in inspection systems, including anti-terrorist activities, industrial diagnostics, new nanotechnologies, and medicine.